\documentclass[%
 reprint,
 amsmath,amssymb,
prb,
]{revtex4-1}
\usepackage{braket}
\usepackage[bookmarksnumbered,colorlinks, linktocpage=true, linkcolor=red, citecolor=blue, filecolor=black,urlcolor=blue]{hyperref}
\usepackage{graphicx}
\usepackage{dcolumn}
\usepackage{bm}

\newcommand{\conffig}[5]{
  \begin{figure*}
      \centering
      \includegraphics[width=#5\textwidth]{#1}
      \caption{\textit{\textbf{#2:}} #3}
      \label{#4}
  \end{figure*}
}

\newcommand{\vhalffig}[5]{
  \begin{figure}
      \centering
      \vspace{-0.65cm}
      \includegraphics[width=#5\textwidth]{#1}
      \caption{\textit{\textbf{#2:}} #3}
      \label{#4}
  \end{figure}
}

\newcommand{\twohalffig}[6]{
  \begin{figure}
      \centering
      \includegraphics[width=#5\textwidth]{#1}
       \includegraphics[width=#5\textwidth]{#6}
      \caption{\textit{\textbf{#2:}} #3}
      \label{#4}
  \end{figure}
}

\begin{document}

\preprint{APS/123-QED}

\title{Superfluidity of strongly correlated bosons in two- and three-dimensional traps}

\author{T. Dornheim$^1$}
 \email{dornheim@theo-physik.uni-kiel.de}
\author{A. Filinov$^{1,2}$}%
\author{M. Bonitz$^1$}
\affiliation{%
$^1$ Institut f\"{u}r Theoretische Physik und Astrophysik, Christian-Albrechts-Universit\"{a}t, Leibnizstra{\ss}e 15, Kiel D-24098, Germany
}%
\affiliation{
$^2$Joint Institute for High Temperatures RAS, Izhorskaya Str. 13, 125412 Moscow, Russia
}



\date{\today}

\begin{abstract}
 We analyze the superfluid crossover of harmonically confined bosons with long-range interaction in both two and three dimensions in a broad parameter range from weak to strong coupling.
We observe that the onset of superfluidity occurs in $3D$ at significantly lower temperatures compared to $2D$. This is demonstrated to be a quantum degeneracy effect. In addition, the spatial distribution
 of superfluidity across the shells of the clusters is investigated. It is found that superfluidity is substantially reduced in the outer layers due to increased correlation effects.
\end{abstract}


\maketitle

\section{\label{sec:level1}Introduction}

Superfluidity (SF), i.e., the emergence of frictionless flow below a critical temperature, 
 is one of the most astonishing manifestations of quantum coherence on a macroscopic scale. It
was first observed in 1938 in liquid 4-He.
In bulk systems, SF is known to being closely related to Bose-Einstein condensation (BEC) and
a consequence of off-diagonal long range order of the density matrix.\citep{yushi}
This concept, obviously, is not directly applicable to finite size systems in traps. Moroever, these systems are possibly strongly inhomogeneous, as a result of boundary or confinement effects. An alternative approach to superfluidity is then to use the probability of realization of exchange cycles involving a sufficiently large number of particles.

A well studied example for such mesoscopic quantum systems are parahydrogen clusters where, for a small number of particles, $N\sim 10$, the superfluid properties explicitly depend on the precise particle number and
the symmetry,\citep{khai} and the superfluidity is essentially uniformly distributed among the system.\citep{mezza,idowu}
In addition, both solid-like order and SF do coexist,\citep{idowu} thus qualifying as a supersolid (see e.g. \citep{supersolid}) state of matter.
Another interesting test case are harmonically confined two-dimensional Coulomb clusters, whose behavior can be controlled by varying the confinement field strength.
There it has been found that not only the global superfluid fraction, but also the spatial distribution of superfluidity crucially depends on $N$.\citep{fili}
In particular, for magic numbers, e.g. $N=12,19,\dots$ (closed shells), there is a strong hexagonal symmetry at the center of the trap and the superfluid density is pushed to the boundary of the system.

While the dependence of superfluidity on temperature, system size and coupling strength is well understood,
the effect of the system dimensionality in finite systems has not been studied systematically. In this work, we try to (partly) fill this gap by studying large spherically confined Coulomb clusters in $2D$ and $3D$
and compare results for, both, the global and spatial distribution of SF. The selected system size $N=150$ is representative 
and large enough to eliminate the sensitivity of the properties to the precise particle number.\cite{ludwig05} Furthermore, our simulations cover the entire range from weak to strong coupling.

The analysis presented in this paper is primarily concerned with fundamental properties of superfluidity in strongly correlated spatially confined bosons in 2D and 3D traps. 
Examples include finite molecular bosonic clusters,\citep{khai,mezza,idowu,kwon,mebo} e.g.\ hydrogen or helium droplets, 
 doublons in strongly correlated solids or ultracold lattice gases \citep{santos,rosch} as well as dipole-interacting trapped gases.\citep{dalf,mybo}
In addition, there are numerous theoretical studies of the fundamental properties of the charged Bose gas (CBG) \citep{koscik,depalo,kim}
and a variety of applications of the model to condensed matter physics.
Indirect excitons \citep{lozovik,timo,sperlich} exhibit, despite their dipole-like interaction at long-range,
a Coulomb-like repulsion at high density.\citep{boning}
Furthermore, the $3D$ CBG of spatially bound electron pairs (small bipolarons) is discussed in the context of high $T_c$ superconductivity
in cuprates.\citep{alexandrov,alexandrov2,lanzara,khasanov}
 Finally, applications also extend to macroscopic 3D bosonic Coulomb systems in the inner regions of neutron stars (proton pairing) \citep{peth,chamel,entrail} and the core of helium white dwarfs.\citep{white,white2}

\section{\label{sec:level11}Model and simulation idea}
We consider two- and three-dimensional systems of $N$ identical bosons in a harmonic confinement potential with frequency $\Omega$ which
are described by the dimensionless Hamiltonian 
\begin{eqnarray}
 \label{hamiltonian}\hat{H} = -\frac{1}{2}\sum_{k=1}^N \nabla_k^2 + \frac{1}{2}\sum_{k=1}^N \mathbf{r}_k^2 + \frac{1}{2} \sum_{k\ne l}^N \frac{\lambda}{|\mathbf{r}_k - \mathbf{r}_l|} \quad ,
\end{eqnarray}
and oscillator units (i.e., characteristic length $l_0=\sqrt{\hbar/m\Omega}$ and energy scale $E_0=\hbar\Omega$) are used throughout this work. 
The selected Coulomb repulsion in Eq.\ (\ref{hamiltonian}) serves as a test-case for a broad class of isotropic long-range interactions and the coupling constant $\lambda = q^2/(\hbar \Omega l_0)$ (with $q$ being the charge) can be controlled experimentally 
by the variation of the trap frequency. 
To simulate the system of interest at a particular inverse temperature $\beta=E_0/k_BT$ in thermodynamic equilibrium, we employ a realization of the widely used worm algorithm path integral Monte-Carlo method.\citep{wa}
The latter delivers quasi-exact results by performing a Trotter decomposition of the density operator $\hat{\rho}=\textnormal{exp}(-\beta\hat{H})$ and each particle is represented by a path of $P$ positions (often denoted as ``beads'' or ``time slices'') in the imaginary time.
To achieve sufficiently well converged results, we typically use $P=80\dots410$ while performing $N_\textnormal{MC}\sim 10^7$ independent measurements.

\section{\label{sec:level111}Results}
\subsection{\label{sec:level2}Superfluid crossover in $2D$ and $3D$}

A convenient way to define superfluidity for finite sized, trapped systems is the consideration of the change of the moment of inertia due to quantum effects, also denoted as \textit{Hess-Fairbank}-effect.
Within the Landau
two-fluid model the total density $n$ is decomposed into a normal and a superfluid component, $n=n_\textnormal{n} + n_\textnormal{sf}$.
Of particular interest is the superfluid fraction $\gamma_\textnormal{sf}$, i.e., the fraction of particles that do not participate in a rotation of the system. 
This quantity can be expressed in terms of the moments of inertia and, within the path integral picture, one can define an estimator as \citep{sf}
\begin{eqnarray}
 \label{sf}\gamma_\textnormal{sf} = \frac{n_\textnormal{sf}}{n} = \frac{I_\textnormal{cl} - I}{I_\textnormal{cl}} = \frac{4m^2\braket{A_z^2}}{\beta\hbar^2I_\textnormal{cl}} \quad ,
\end{eqnarray}
with the total and classical moment of inertia $I$ and $I_\textnormal{cl}$, respectively, and the area $\mathbf{A}$ which is enclosed by the particle trajectories
\begin{eqnarray}
 \label{area}   \mathbf{A} = \frac{1}{2}\sum_{k=1}^N\sum_{i=1}^{P} \left(\mathbf{r}_{k,i} \times \mathbf{r}_{k,i+1}\right) \quad .
\end{eqnarray}
In Eq.\ (\ref{sf}) the system is assumed to be set into rotation around the $z$-axis and, hence, only the area in the $x$-$y$-plane $A_z$ is relevant.

   \begin{figure}
      \centering
      \includegraphics[width=0.48\textwidth]{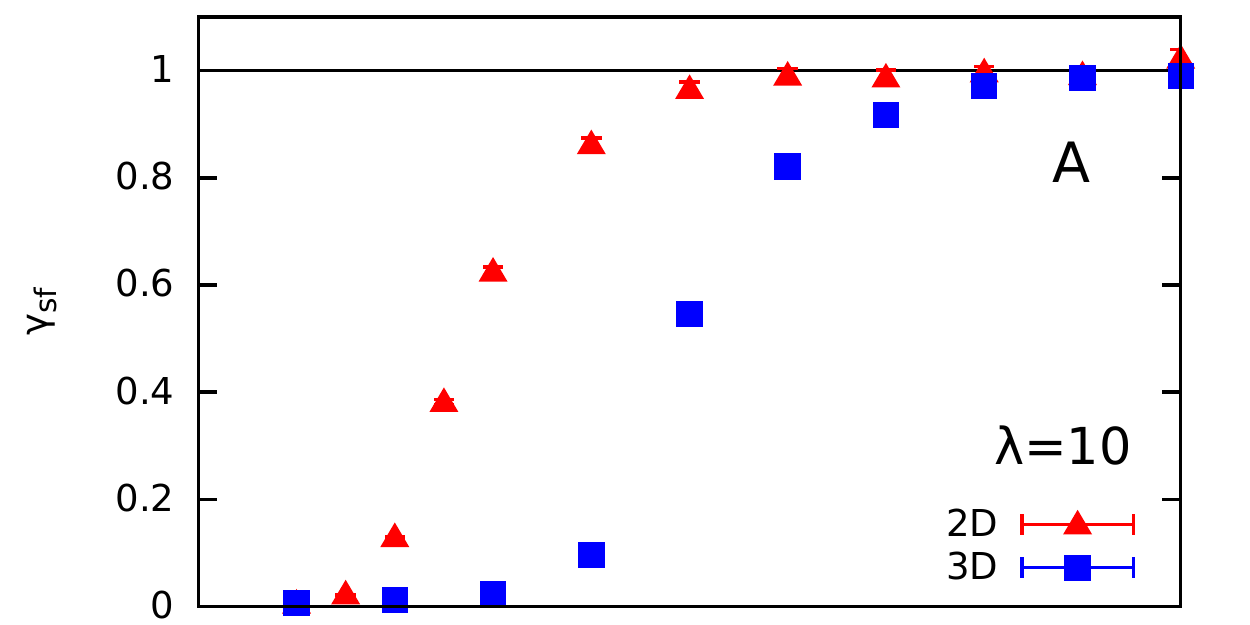}\\
      \vspace{-0.15cm}
       \includegraphics[width=0.48\textwidth]{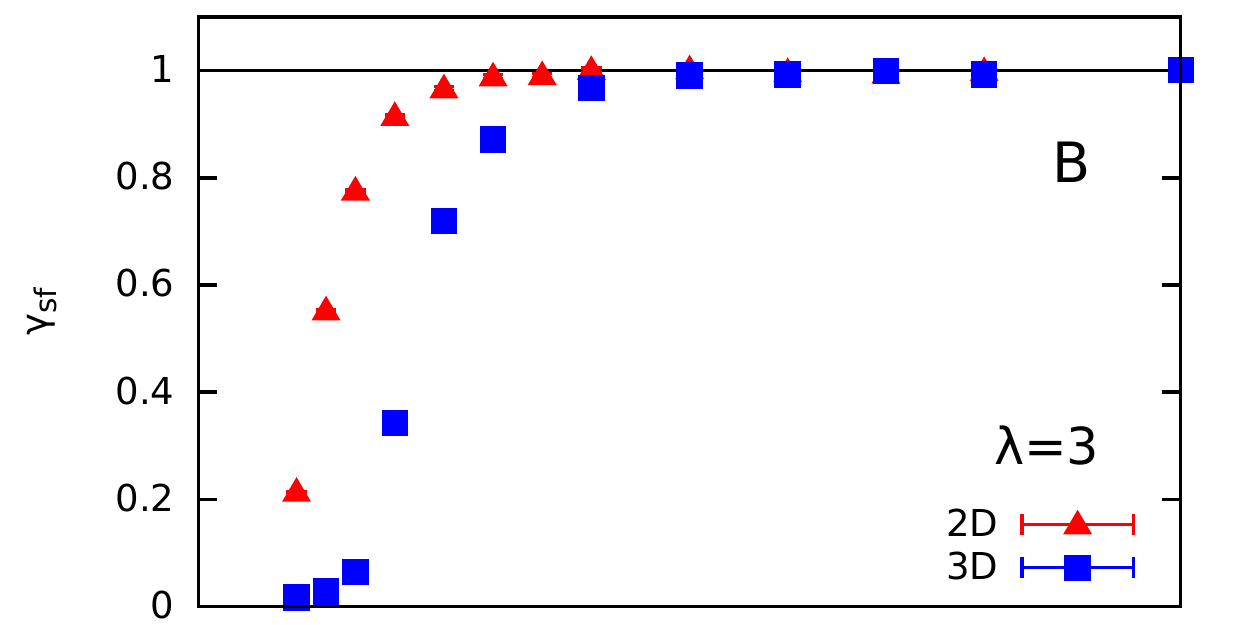}\\
       \vspace{-1.cm}
       \includegraphics[width=0.48\textwidth]{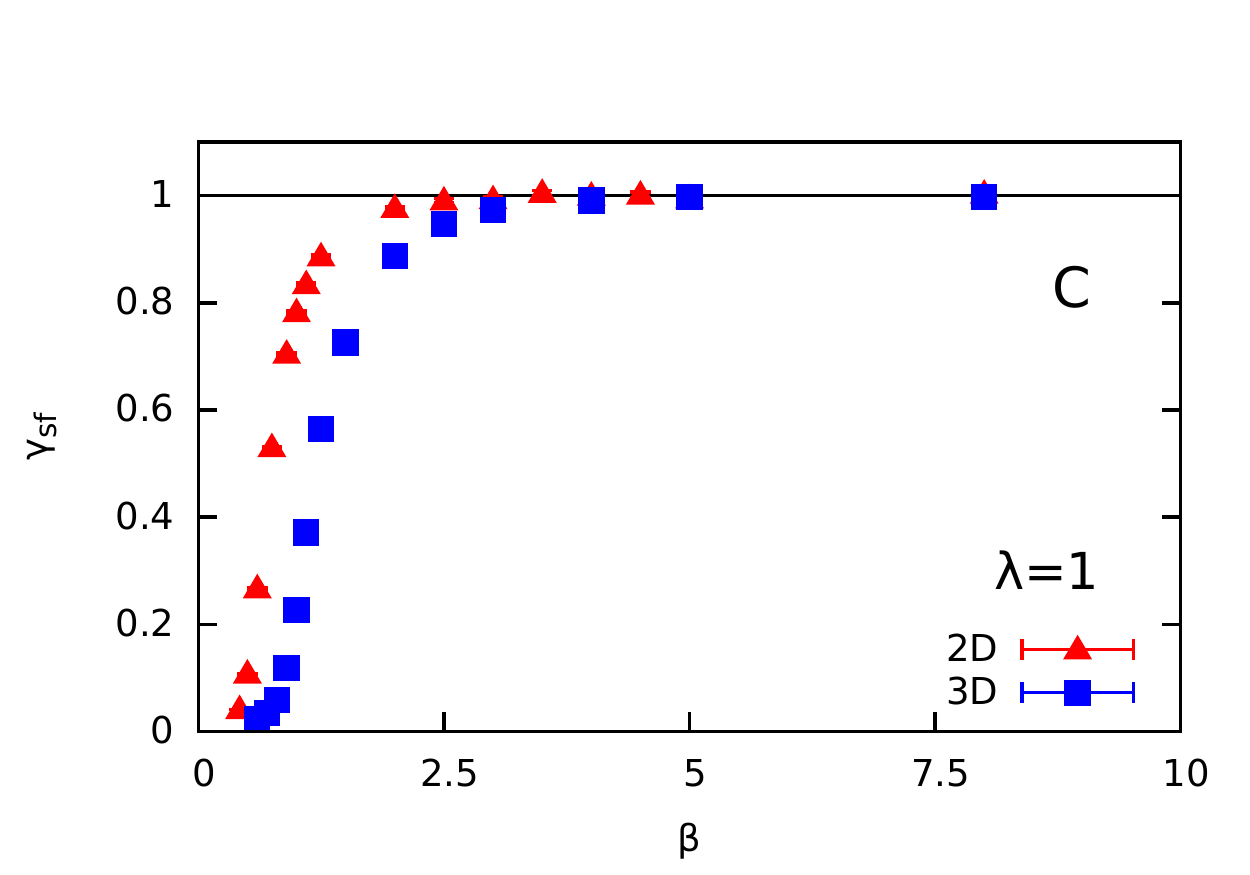}
      \caption{\textit{\textbf{Superfluid crossover in $2D$ and $3D$:}} The superfluid fraction $\gamma_\textnormal{sf}$ is plotted over the inverse
  temperature $\beta$ for $N=150$ particles in $2D$ (red triangles) and $3D$ (blue squares) for the coupling constant $\lambda=10$ (A), $\lambda=3$ (B) and $\lambda=1$ (C).}
      \label{picture}
  \end{figure}
  
The results for the superfluid fraction, Eq.\ (\ref{sf}), are shown in Fig.\ \ref{picture}, where $\gamma_\textnormal{sf}$ is plotted versus the inverse temperature $\beta$ for $N=150$ particles
in $2D$ (red triangles) and $3D$ (blue squares) and for the coupling parameters $\lambda=10$ (A), $\lambda=3$ (B) and $\lambda=1$ (C). All curves exhibit the expected increase of superfluidity with increasing $\beta$ and saturate to unity, i.e., a completely superfluid system. In addition, the crossover \cite{phase_trans_finite} is shifted to lower temperature with increasing particle interaction. This is a direct consequence of the increased inter-particle distance, which reduces the probability of the occurence of exchange cycles. However, it is interesting to note that, for all $\lambda$, the onset of the crossover occurs at lower temperature for the $3D$ system.

This is a non-trivial observation and could be caused by different effects: 
(i) The availability of the additional dimension leads to a reduced degeneracy. This decreases exchange effects and hence, explains the observed shift of the onset superfluidity.
(ii) The three-dimensional nature of the particle exchange could lead to a reduction of the projection of the area-vector $\mathbf{A}$ from Eq.\ (\ref{area})
onto a particular plane. In this case, one would observe the onset of the crossover at an increased degeneracy compared to the $2D$ system with the same parameters.
To decide which (if any) of the two explanations is most likely we analyze the probability $P(L)$ for a single bead to be involved in an exchange cycle
which consists of $L$ particles and, in addition, a degeneracy parameter $\chi$ [cf. Eq.\ (\ref{chi})].
For (i) to be correct, we expect similar values of $\chi$ and $P(L)$, for the same amount of superfluidity in the system. For (ii), on the other hand, we would expect a significantly
increased $\chi$ and higher probabilities $P(L)$ for $L>1$, for equal $\gamma_\textnormal{sf}$ in $3D$.

In Fig.\ \ref{picture_prob} the product $P(L)L$ (with $\sum_L P(L)L = 1$) is plotted versus $L$ again for $N=150$ particles and $\lambda=3$ for both $2D$ (red) and $3D$ (blue). 
The top image corresponds to an equal inverse temperature $\beta=2$, i.e., two systems with a non-zero and non-unity superfluid fraction $\gamma_\textnormal{sf}(2D)\approx0.92$ and $\gamma_\textnormal{sf}(3D)\approx0.34$.
Both curves exhibit a similar decay with increasing $L$ and a sharp bend for very large exchange cycles, which occurs for smaller particle numbers
in $3D$. In the $3D$ system both single and two particle trajectories are more probable than in $2D$ and the two curves intersect at $L=3$.
All larger exchange cycles are significantly more probable in $2D$.
The bottom image of Fig.\ \ref{picture_prob} shows the same information but for a fixed superfluid fraction $\gamma_\textnormal{sf}\approx0.6$.
Here, the probability to be involved in a particular exchange cycle is nearly equal for both dimensionalities and the two curves intersect several times.
In addition, the sharp bend for large $L$ occurs at the same position.
This is a first hint towards explanation (i) because the same amount of superfluidity requires comparable realization rates of exchange cycles
while, at the same inverse temperature, exchange is suppressed in $3D$ which explains the later onset of superfluidity compared to $2D$. For completeness, we note that, both $\lambda=1$ and $\lambda=10$, exhibit the same behavior as shown for $\lambda=3$ in Fig.\ \ref{picture_prob}.

 \vhalffig{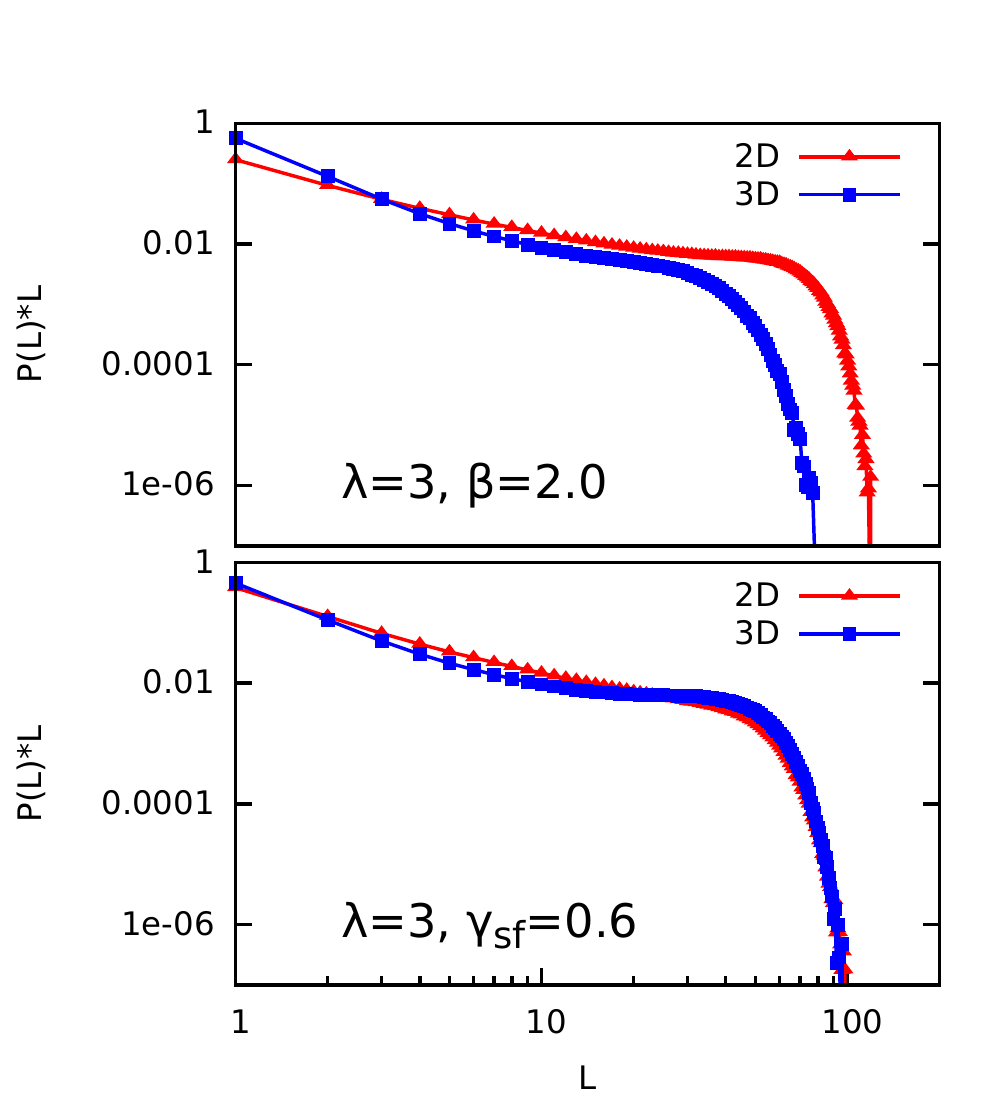}{Probability distribution of exchange cycles in $2D$ and $3D$}{The probability for a single bead to be involved in an exchange cycle of length $L$, $P(L)L$, is plotted over the former for $N=150$ particles and $\lambda=3$. The top image corresponds to $\beta=2$ with $\gamma_\textnormal{sf}(2D)\approx0.92$ and $\gamma_\textnormal{sf}(3D)\approx0.34$ and for the bottom image the inverse temperature has been adjusted that $\gamma_\textnormal{sf}\approx0.6$ in both dimensionalities, i.e., $\beta(2D)=1.35$ and $\beta(3D)=2.27$.}{picture_prob}{0.48}

Next, we define the degeneracy parameter as
\begin{eqnarray}
 \label{chi} \chi = \overline{n}\lambda_\beta^d \quad ,
\end{eqnarray}
with the dimensionality $d$, the thermal de-Broglie wavelength $\lambda_\beta=\hbar \sqrt{2\pi\beta/m}$ and the mean density $\overline{n}$ (in $2D$ and $3D$, we compute the radial density $n$ by averaging over angular and spherical segments, respectively, with a fixed distance from the trap center), which has been averaged over the radial extension of the particular system.
Thus, Eq.\ (\ref{chi}) provides a measure for the average number of particles within the approximate extension of a 
single-particle wavefunction.
It should be noted that the strong inhomogeneity of the density of correlated trapped quantum particles, cf. Fig.\ \ref{picture3}, makes the definition in Eq.\ (\ref{chi}) arbitrary to some degree.
However, we found that using $\overline{n}$ gives reliable results even for different shell structures.

\twohalffig{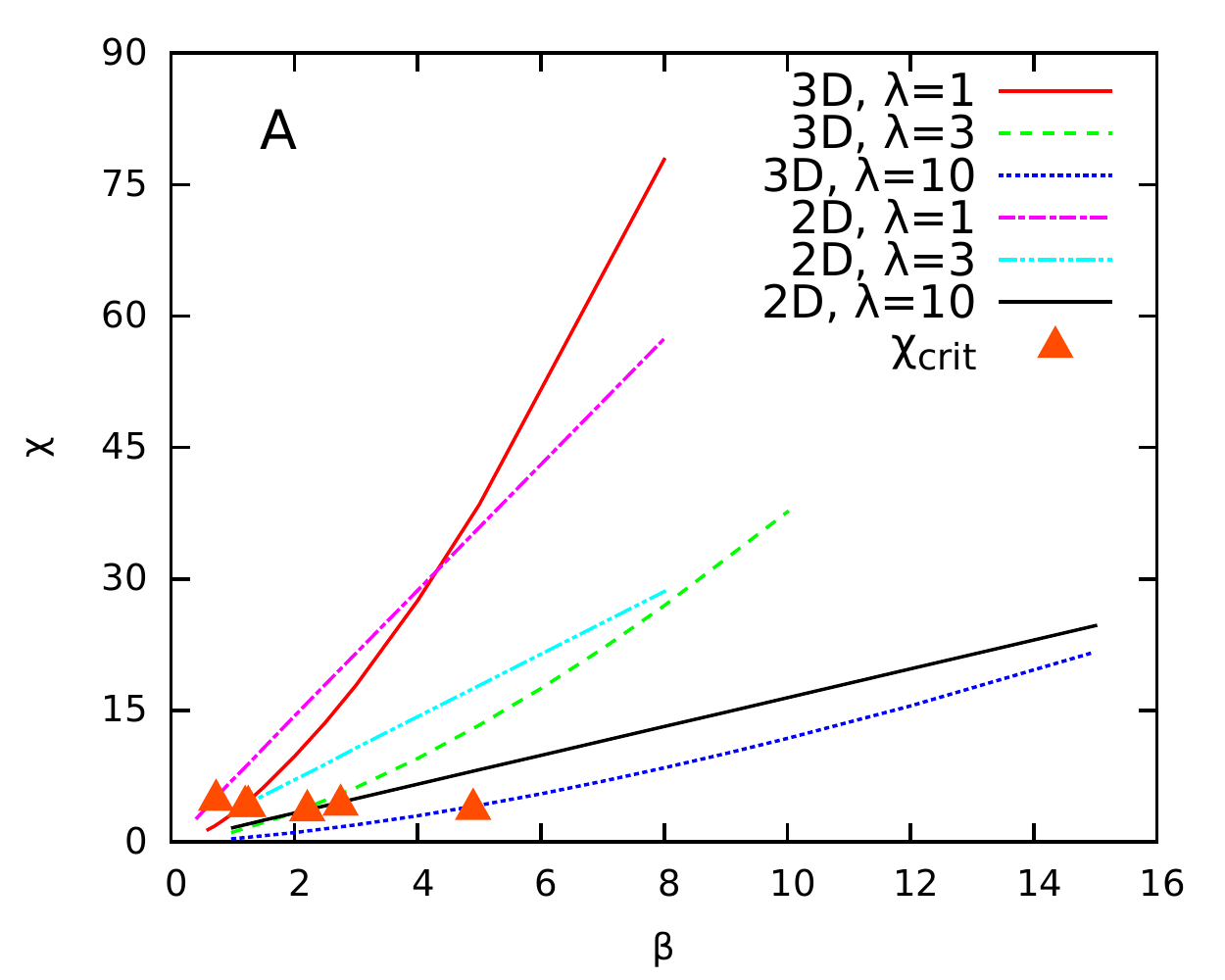}{Degeneracy parameter}{The degeneracy parameter $\chi$ is plotted over the inverse
  temperature $\beta$ for the systems from Fig.\ \ref{picture}. The critical value $\chi_\textnormal{crit}$ (orange triangles) marks the superfluid crossover, i.e., $\gamma_\textnormal{sf}=0.5$.
   Image A covers the entire inverse temperature range of our simulations and image B shows a magnified segment around the superfluid crossover.}{picture2}{0.48}{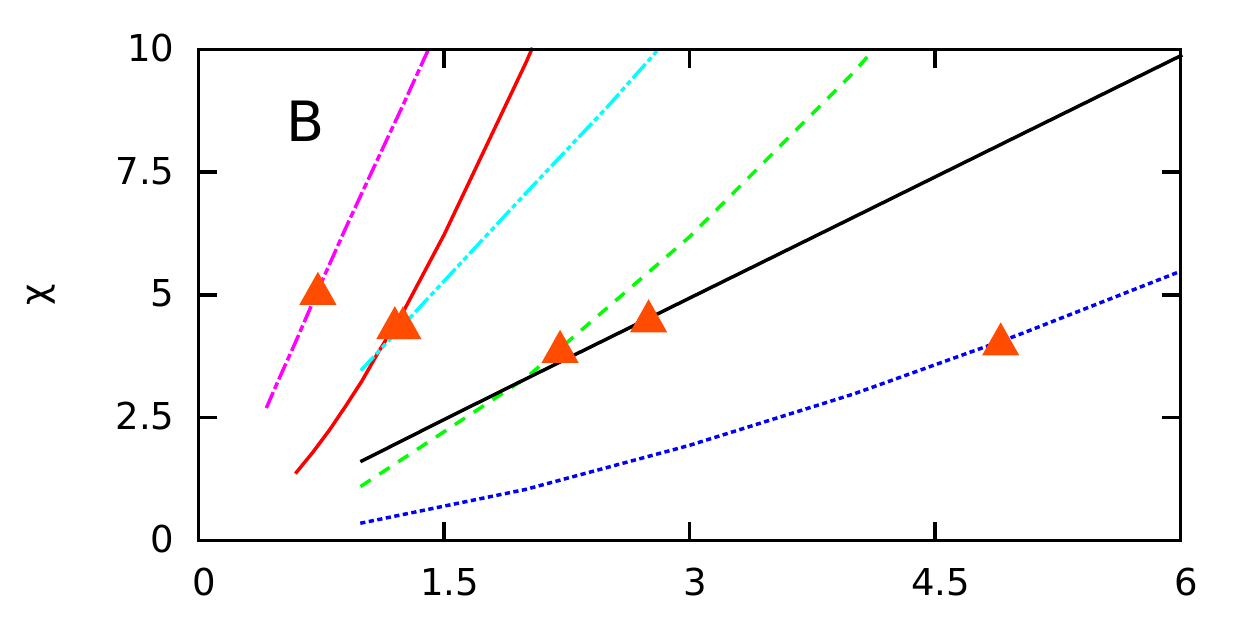}

The results for Eq.\ (\ref{chi}) are shown in Fig.\ \ref{picture2} (A) where $\chi$ is plotted versus $\beta$ for the systems from Fig.\ \ref{picture}.
The degeneracy increases with $\beta$ for all simulated systems, as expected. 
However, a comparison between $2D$ and $3D$ systems with otherwise equal parameters reveals that, at high $T$, the three-dimensional system exhibits a smaller $\chi$, until the two curves intersect. This is an immediate consequence of the definition \ (\ref{chi}), where the de-Broglie wavelength enters with the power of the dimensionality $d$. 
This implies that, for a constant mean density $\overline{n}$, which becomes valid with increasing $\beta$, it is $\chi(d=2)\propto\beta$ and $\chi(d=3)\propto\beta^{3/2}$, i.e., a linear and polynomial behavior with respect to the inverse temperature in $2D$ and $3D$, respectively.
The connection between Fig.\ \ref{picture2} and the superfluid crossover from Fig.\ \ref{picture} is given by the orange triangles, which mark the critical degeneracy $\chi_\textnormal{crit}$, i.e., the degeneracy parameter at the inverse temperature for which $\gamma_\textnormal{sf}=0.5$.
Our simulations have revealed that, despite the onset of superfluidity at significantly lower temperature in $3D$, $\chi_\textnormal{crit}$ takes similar values for both dimensionalities.
In fact, the $2D$ systems from Fig.\ \ref{picture2} even exhibit a slightly higher $\chi$ for the critical superfluid fraction than their $3D$ counterparts, cf.\ Fig.\ \ref{picture2} (B), where a magnified segment around the crossover is shown.
However, this rather peculiar feature should not be over-interpretated due to the average character of the definition of the degeneracy parameter itself.
Thus, we conclude that the behavior observed in Fig.\ \ref{picture2} rules out explanation (ii), and the different critical temperatures for the superfluid crossover in $2D$ and $3D$ appear to be a degeneracy effect.

\subsection{Local superfluid density\label{seclsf}}

Another interesting question is how the superfluidity is distributed across the system in the vicinity of the crossover. 
To investigate this topic we use a spatially resolved superfluid density estimator of Kwon \textit{et al.},\citep{kwon} that is consistent with the two-fluid interpretation:
\begin{eqnarray}
\label{nsf} n_\textnormal{sf}(\mathbf{r}) = \frac{4m^2}{\beta\hbar^2 I_\textnormal{cl}}\braket{A_zA_{z,\textnormal{loc}}(\mathbf{r})} \quad .
\end{eqnarray}
Here, $A_{z,\textnormal{loc}}(\mathbf{r})$ denotes a local contribution to the total area enclosed by the particle paths.
In $2D$, the system is assumed to rotate around the axis perpendicular to the trap and the quantity from Eq.\ (\ref{nsf}) can be radially averaged without loosing information.
For a $3D$ trap, however, the axis of rotation causes a break of the spherical symmetry and the definition of a meaningful average is less obvious.
In this work, we average $n_\textnormal{sf}$ over spherical segments with the distance $r$ from the center of the trap.
This gives a quantity that approaches the total radial density $n$ for a completely superfluid system, i.e., for $\gamma_\textnormal{sf}=1$.
Nevertheless, one should keep in mind that, in general, particles with the same distance from the center $|{\bf r}|$ have different contributions to the classical moment of inertia, depending on the orientation of their radius vector ${\bf r}$ with respect to $\mathbf{e}_z$.

\vhalffig{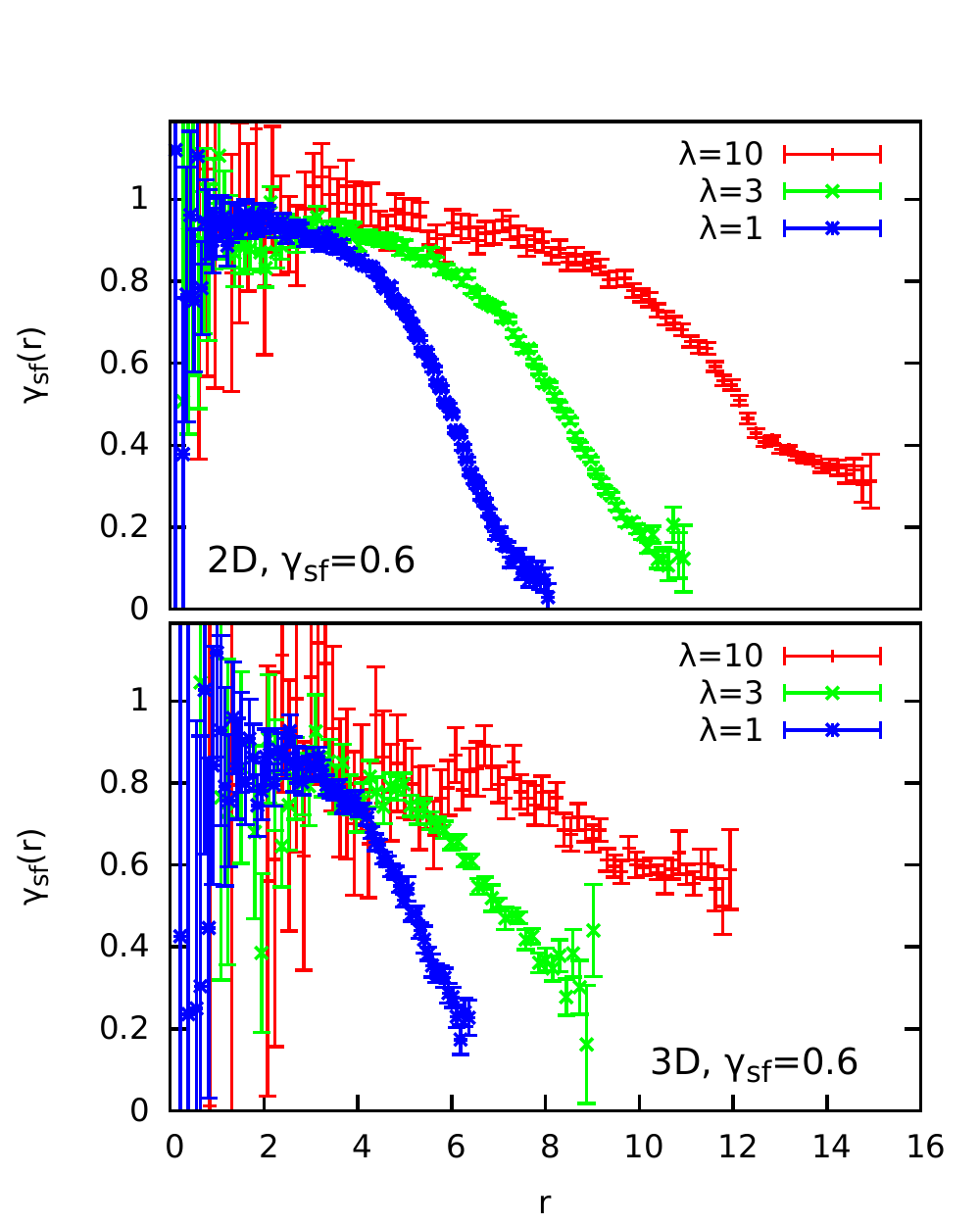}{Local superfluid fraction in $2D$ and $3D$}{The spatially resolved ratio of the superfluid and total density $\gamma_\textnormal{sf}(r)=n_\textnormal{sf}(r)/n(r)$ is plotted over the distance to the center of the trap $r$ for $N=150$ particles in $2D$ (top) and $3D$ (bottom) for a fixed superfluid fraction $\gamma_\textnormal{sf}\approx0.6$. The larger errors around the center of the trap are due to the worse statistics in small angular and spherical segments in $2D$ and $3D$, respectively.}{lsfraction}{0.48}

\conffig{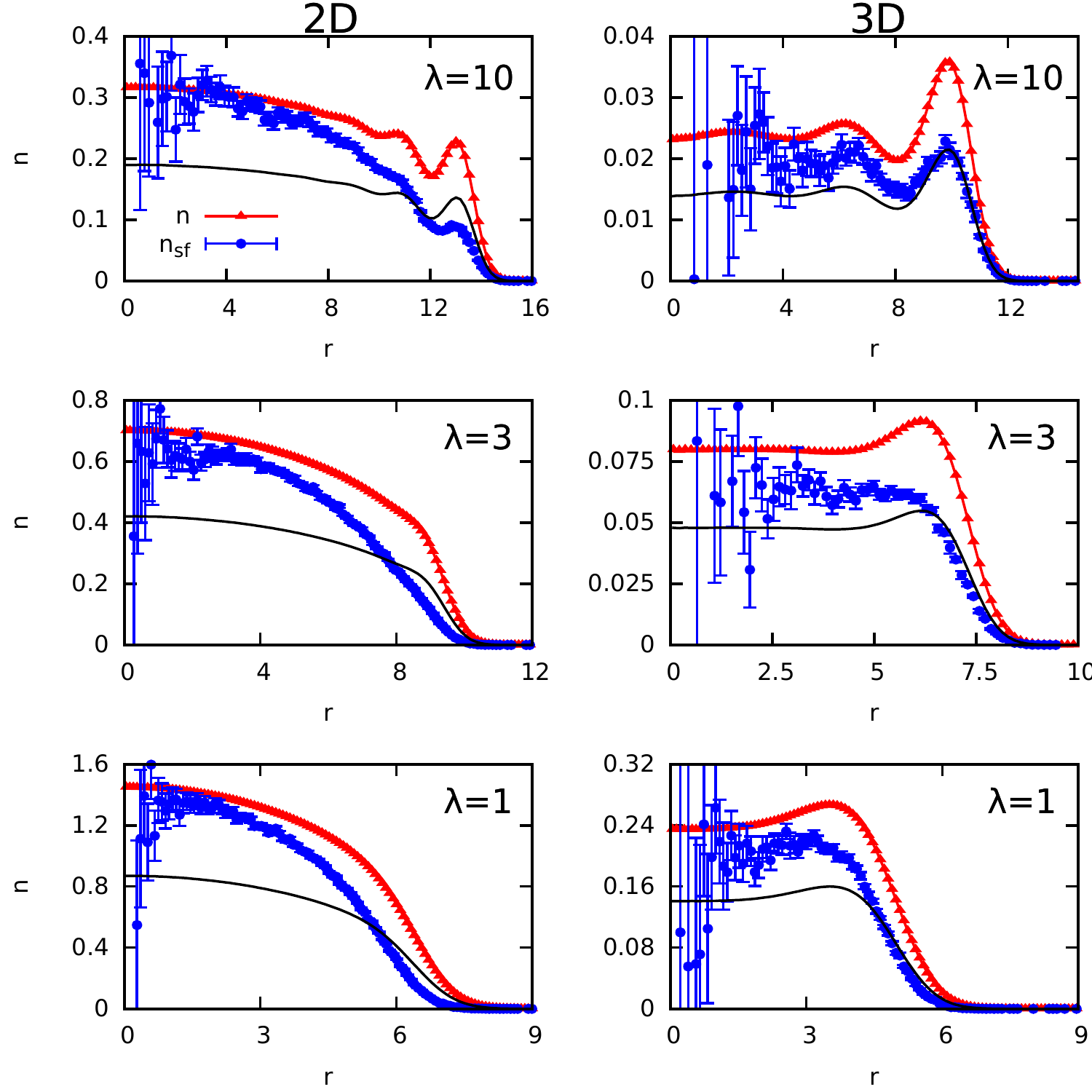}{Spatially resolved superfluidity}{The radial density $n$ (red triangles) and superfluid density $n_\textnormal{sf}$ (blue squares) are plotted with respect to the distance to the center of the trap $r$ for a fixed total superfluid fraction $\gamma_\textnormal{sf}\approx0.6$ for the systems from Figs.\ \ref{picture} and \ref{picture2}. The black curves correspond to a constant local SF ratio $\gamma_\textnormal{sf}(r)=0.6$.}{picture3}{0.90}

Fig.\ \ref{lsfraction} shows the local ratio $\gamma_\textnormal{sf}(r)=n_\textnormal{sf}(r)/n(r)$ for $2D$ (top) and $3D$ (bottom) systems and an
inverse temperature that has been choosen in a way that $\gamma_\textnormal{sf}\approx0.6$.\citep{adjustgamma}
For both dimensionalities $\gamma_\textnormal{sf}(r)$ exhibits a decay with increasing distance from the center of the trap, $r$, which is most distinct
for smaller coupling. This behavior can be understood by considering the normalization of the local superfluid density

\begin{eqnarray}
 \int d\mathbf{r}\ n_\textnormal{sf}(\mathbf{r}) \mathbf{r}_\perp^2 = \gamma_\textnormal{sf} I_\textnormal{cl} \qquad . \label{nsfnorm}
\end{eqnarray}
Eq.\ (\ref{nsfnorm}) implies that $n_\textnormal{sf}$ integrates to the quantum correction to the moment of inertia, i.e., the missing contribution due to 
the particles in the superfluid phase.
However, this also means that 
\begin{eqnarray}
  \frac{1}{N}\int d\mathbf{r}\ n_\textnormal{sf}(\mathbf{r}) \ne \gamma_\textnormal{sf} \quad ,
\end{eqnarray}
because, for the global SF fraction, not only the ratio of superfluid and total particle numbers but, in addition, the position of a particular particle
is relevant. According to Eq.~(\ref{nsfnorm}), particles near the center of the trap have a small contribution to the total moment of inertia and so their superfluidity
only slightly influences $\gamma_\textnormal{sf}$. At the boundary, however, each particle has a significant contribution to $I$ 
and, thus, $n_\textnormal{sf}(r)$ in this region crucially influences the global SF fraction.
With increasing $\lambda$, the interparticle repusion grows and the system extends radially (cf.\ Fig.\ \ref{picture3}), so this effect becomes even more important. For a fixed $\gamma_\textnormal{sf}$, as it is the case in Fig.\ \ref{lsfraction}, the spatially resolved SF fraction in the outermost shell must approach the global value, with increasing coupling strength.
A similar effect has recently been observed by Kulchytskyy \textit{et al.} \citep{kulch} from PIMC simulation of $^4$He in a cylindrical nanopore where the local superfluid response has been obtained, both, with the area formula, Eq.\ (\ref{nsf}), and a winding number approach \citep{khai2} for a rotation around and the flow along the pore axis, respectively.
Despite a very similar spatial distribution for the two estimators, there occurs a significant difference in the global SF fraction due to the different normalizations.

The comparison between the $2D$ and the $3D$ results in Fig.\ \ref{lsfraction} reveals that, for $d=3$ and equal $\lambda$, the local superfluid fraction in the outer region exceeds its two dimensional counterpart. In fact, the uppermost (red) curve ($\lambda=10$) in the bottom image appears to be above the global value \citep{adjustgamma} almost everywhere, 
which seems to violate Eq.\ (\ref{nsfnorm}).
However, this is a consequence of the symmetry break due to the external rotation around $\mathbf{e}_z$ and the applied averaging over spherical segments in $3D$.
For this reason, all particles in the outermost shell have different contributions to the moment of inertia. The expected decrease of $\gamma_\textnormal{sf}(r)$
for the largest $\mathbf{r}_\perp$ is therefore masked in Fig.\ \ref{lsfraction} by contributions near $\mathbf{e}_z$ from the same shell.

The remaining open question is why $\gamma_\textnormal{sf}$ is distributed inhomogeneously in the first place, in particular, why there is more superfluidity around the center of the trap than in the outer regions.
In Fig.\ \ref{picture3}, both the total (red triangles) and superfluid density (blue squares) are plotted with respect to $r$ for the systems from Fig.\ \ref{lsfraction}.
The top row shows density profiles for rather strong coupling, $\lambda=10$, for both the two- (left) and three-dimensional (right) system.
In $2D$, the total density $n$ exhibits a relatively smooth decay with increasing $r$ until, at the boundary of the system, two shell-like oscillations appear. In $3D$, $n$ stays almost constant around the center and a very pronounced shell appears at the boundary. The center and bottom rows show the same information for medium ($\lambda=3$) and weak ($\lambda=1$) coupling, respectively. One clearly sees that the shell-like features disappear
for both dimensionalities when the coupling is reduced. A general feature for all couplings is that, $n$ always decays in 2D whereas, in 3D, it is nearly constant in the center. Also the peak(s) at the boundary are always significantly stronger in three dimensions.

The local superfluid density, on the other hand, exhibits a rather surprising behavior.
For $d=2$ and strong coupling, $n_\textnormal{sf}$ equals $n$, around the center of the trap, whereas it is significantly suppressed for larger $r$, in particular within the
outermost shell. The $3D$ counterpart exhibits a similar behavior. The two innermost, weakly pronounced shells are nearly completely superfluid, whereas the difference between $n$ and $n_\textnormal{sf}$ is the largest in the outermost shell.
This non-uniformity of the superfluidity can be made more transparent by comparing to the case of a constant local superfluid fraction $\gamma_\textnormal{sf}(r)=0.6$ which is depicted by the black curves in Fig.\ \ref{picture3}. Clearly, the true superfluidity is always above (below) the black curve in the center (at the edge).

Let us discuss the origin of this spatial variation of the superfluid density. The first reason that comes to mind is the spatial variation 
of the local degeneracy parameter which is proportional to the local total density,  $\chi(r)\sim n(r)$.
However, the numerical results for $n_\textnormal{sf}$ clearly deviate from this suggestion. In particular, for $\lambda=1, 3$ the superfluid density does not follow 
the local degeneracy (total density). Only at $\lambda=10$ when shells are formed,  $n_\textnormal{sf}$ increases around the outermost shell. However, the absolute value of the superfluid density does not reach a maximum despite the maximum of the degeneracy parameter.
From this we conclude that the spatial distribution of superfluidity is not governed by the spatial variation of degeneracy but by the local strength of correlations.

In fact, it is well known that spatial order leads to a reduction of superfluidity. 
In Coulomb systems, particles mainly experience interaction with neighbors with smaller $r$, whereas most of the pair interactions with particles from the outer parts of the system cancel. This is a pure screening effect (Faraday cage effect), and cancellation would be complete in mean field approximation. Hence, the investigated Coulomb clusters clearly exhibit the strongest order around the boundary and are less correlated around the center of the trap. This is also manifest in the shell formation at the outer boundary (red curves in Fig.\ \ref{picture3}) which is observed in classical Coulomb clusters as well.\citep{hanno} Another mechanism that suppresses superfluidity at the cluster boundary is a geometrical effect: within the outermost shell there are fewer particles available for the formation of exchange cycles due to the lack of neighbors in radial direction.

The medium and weakly coupled system in $2D$ exhibit the same trend as for $\lambda=10$.
In $3D$, the largest deviation between $n$ and $n_\textnormal{sf}$ appears around the outermost shell as well, whereas the superfluid density remains nearly constant over the rest of the system. This is in contrast to $2D$ and arises from the different behaviors of the total density $n(r)$. At the same time, the spatially resolved SF fraction (cf.\ Fig.\ \ref{lsfraction}) exhibits similar trends in $2D$ and $3D$.

\section{Discussion}
In summary, we have presented ab-initio results for the superfluid properties of Coulomb interacting bosons in both $2D$ and $3D$ harmonic traps.
It was revealed that the availability of an additional dimension causes a decrease of the critical temperature of the superfluid crossover.
To explain this non-trivial feature, we have analyzed the probability distribution of exchange cycles and the degeneracy parameter (\ref{chi}), and it was revealed that the critical superfluid fraction, $\gamma_\textnormal{sf}=0.5$,
occurs for similar degeneracy for both dimensionalities,
despite the significantly higher $\beta$ in $3D$.
In addition, we have investigated the spatial distribution of superfluidity across the system by using a local superfluid density estimator.
In both $2D$ and $3D$, the largest difference between the total and superfluid density $n$ and $n_\textnormal{sf}$, occurs near the boundary
of the system. This is a direct consequence of the increased order at large distances from the center of the trap.

The onset of superfluidity at comparatively lower temperature in $3D$ systems is expected to be a general feature and not specific for Coulomb interaction.
For the investigated particle number $N=150$, superfluidity requires a collective response of the entire system, in particular the frequent realization of large exchange cycles in the path integral picture. The probability for the latter crucially depends on the degeneracy.
The reported $\beta$-dependence of the degeneracy parameter $\chi$ is valid for other long-range interactions as well, because the coupling only enters 
in the average density $\overline{n}$ which remains constant with decreasing $T$.

The observed non-uniform distribution of superfluidity in the vicinity of the crossover, on the other hand, depends on several system properties.
For small $N$ and strong coupling, the exact particle number plays an important role and, in case of strong hexagonal order at the center of the trap,
the superfluidity is essentially located at the boundary.\citep{fili}
The almost complete screening of the interaction of a given particle with particles located at a larger distance from the center is a Coulomb specific effect. In the case of other pair interactions, a force from the outer regions may exist, which leads to different density profiles and spatial distributions of correlation effects.\citep{hanno}
In addition, the explicit choice of the confinement potential also influences these quantities. The application of e.g. a
quartic trap is expected to enhance the  behavior observed in Sec.\ (\ref{seclsf}), whereas a weaker than harmonic confinement is expected to reduce these trends.

While we expect that the observed trends are typical for Bose systems in confinement potentials,
we mention that, for unconfined finite systems with attractive interaction, different behaviors have been predicted.
Khairallah \textit{et al.} reported that, for mesoscopic parahydrogen clusters, superfluidity could be realized by loosely bound surface molecules,\citep{khai} although this is in disagreement with results by Mezzacapo and Boninsegni.\citep{mezza}

\section*{\small Acknowledgements}
This work is supported by the Deutsche Forschungsgemeinschaft via SFB-TR24 and project FI 1252/2 and by grant SMP006 for CPU time at the HLRN.


\begin{thebibliography}{23}
\bibitem{phase_trans_finite} While the notion ``phase transition'' refers to infinite systems very similar structural transitions are observed in finite systems in traps \cite{filinov_prl01}. The main difference is that the transition region is broadened. Therefore, we will use the term ``crossover'' here.
\bibitem{adjustgamma} For each simulation the input parameters $\beta$ and $\lambda$ are given, which allows to achieve the target value $\gamma_\textnormal{sf}=0.6$ only approximately. 
For the six cases shown in Figs.\ \ref{lsfraction} and \ref{picture3} we calculated the global SF fractions 
\begin{center}
    \begin{tabular}{ c  c  c  c }
     & & & \vspace{-0.1cm}\\ 
    \hline  \hline 
    & \vspace{-0.1cm}& & \vspace{-0.2cm} \\
   $\lambda$ & $1$ & $3$ & $10$ \\  \hline  
  & & & \vspace{-0.2cm} \\
  \hspace{0.3cm} $2D$: & \hspace{0.3cm}$0.59(9)$\hspace{0.3cm} & \hspace{0.3cm}$0.59(8)$\hspace{0.3cm} &\hspace{0.3cm} $0.62(1)$\hspace{0.3cm}\vspace{0.1cm} \\
   \hspace{0.3cm} $3D$: & $0.59(0)$ & $0.58(8)$ & \hspace{0.21cm}$0.64(7)$  \vspace{0.08cm}\\ \hline \hline
   & & & \vspace{-0.1cm}
    \end{tabular}
   \end{center}
The obtained $\gamma_\textnormal{sf}$ are reasonably close to the target, although, for $\lambda=10$ and $3D$, a rather significant difference occurs.
However, the deviations from the target value are smaller than the statistical uncertainty of $\gamma_\textnormal{sf}(r)$ and a reliable comparison between the investigated $\lambda$ and the $2D$ and $3D$ systems is possible.
\end{thebibliography}
\end{document}